\documentclass[sigconf, nonacm]{acmart}
\setcopyright{none}
\renewcommand\footnotetextcopyrightpermission[1]{}

\usepackage{fancyhdr}
\fancypagestyle{plain}{%
   \fancyhf{} %
   \fancyfoot[L]{ACM SIGCOMM Computer Communication Review}%
   \fancyfoot[R]{Volume 57 Issue 1, January 2025}%
}
\fancypagestyle{firstpagestyle}{%
   \fancyhf{} %
   \fancyfoot[L]{ACM SIGCOMM Computer Communication Review}%
   \fancyfoot[R]{Volume 57 Issue 1, January 2025}%
}

\usepackage[utf8]{inputenc}
\usepackage[T1]{fontenc}
\usepackage[english]{babel}

\usepackage{amsfonts}
\usepackage{mathtools}
\usepackage{balance}
\usepackage{url}
\usepackage{caption}
\usepackage{multirow}
\usepackage{booktabs}
\usepackage{xcolor}
\usepackage{numprint}

\usepackage{amsmath,amsfonts}
\usepackage{algorithmic}
\usepackage{graphicx}
\usepackage{textcomp}
\usepackage{xcolor}
\usepackage{tabulary}
\usepackage{comment}
\usepackage{siunitx} 
\usepackage{svg}
\usepackage{pifont}
\usepackage{enumitem}
\usepackage{censor}
\usepackage{flushend}
\usepackage{xspace}
\usepackage{xcolor,colortbl}

\usepackage[acronym, nomain, nowarn, automake]{glossaries}

\newcommand\todo[1]{{\color{red} [TODO: #1]}}
\newcommand{\sge}[1]{\textbf{\textcolor{orange}{\footnotesize\textsuperscript{[SG]} #1}}} 

\newif\ifcomment
\commentfalse

\ifcomment
    \newcounter{MVNumberOfComments}
    \newcounter{AELNumberOfComments}
    \newcounter{VVNumberOfComments}
    \stepcounter{MVNumberOfComments}
    \stepcounter{AELNumberOfComments}
    \stepcounter{VVNumberOfComments}
    \newcommand{\mvnote}[1]{\textcolor{blue}{\small \bf [MV\#\arabic{MVNumberOfComments}\stepcounter{MVNumberOfComments}: #1]}}

\else
    \newcommand\mvnote[1]{}
\fi

\ifcomment
    \newcounter{DJNumberOfComments}
    \stepcounter{DJNumberOfComments}
    \newcommand{\djnote}[1]{\textcolor{green}{\small \bf [daniel\#\arabic{DJNumberOfComments}\stepcounter{DJNumberOfComments}: #1]}}    
\else
    \newcommand\djnote[1]{}
\fi

\newcommand{\ie}{{\it i.e.,}\xspace}
\newcommand{\eg}{{e.g.,}\xspace}

\makeglossaries 

\newacronym{mvno}{MVNO}{Mobile Virtual Network Operator}
\newacronym{iot}{IoT}{Internet of Things}
\newacronym{wlan}{WLAN}{Wireless Local Area Network}
\newacronym[shortplural=ASes, longplural=Autonomous Systems (ASes)]{as}{AS}{Autonomous System}
\newacronym{IGP}{IGP}{Interior Gateway Protocol}
\newacronym{egp}{EGP}{Exterior Gateway Protocol}
\newacronym{tcp}{TCP}{Transmission Control Protocol}
\newacronym{bgp}{BGP}{Border Gateway Protocol}
\newacronym{ip}{IP}{Internet Protocol}
\newacronym{ipx}{IPX}{IP Packet eXchange}
\newacronym{mno}{MNO}{Mobile Network Operator}
\newacronym{gtp}{GTP}{GPRS Tunneling Protocol}
\newacronym{cidr}{CIDR}{Classless Inter-Domain Routing}
\newacronym{asn}{ASN}{\gls{as} Number}
\newacronym{ir}{IR}{Internet Registry}
\newacronym{iana}{IANA}{Internet Assigned Numbers Authority}
\newacronym{icann}{ICANN}{Internet Corporation for Assigned Names and Numbers}
\newacronym{rir}{RIRs}{Regional Internet Registries}
\newacronym{lir}{LIRs}{Local Internet Registries}
\newacronym{roa}{ROA}{Route Origin Attestation}
\newacronym{ebgp}{eBGP}{external BGP}
\newacronym{ibgp}{iBGP}{internal BGP}
\newacronym{nlri}{NLRI}{Network Layer Reachability Information}
\newacronym{amps}{AMPS}{Advanced Mobile Phone System}
\newacronym{gsm}{GSM}{Global System for Mobile Communication}
\newacronym{bts}{BTS}{Base Transceiver Station}
\newacronym{ms}{MS}{Mobile Station}
\newacronym{bsc}{BSC}{Base Station Controller}
\newacronym{bss}{BSS}{Basestation Subsystem}
\newacronym{ran}{RAN}{Radio Access Network}
\newacronym{msc}{MSC}{Mobile Switching Center}
\newacronym{vlr}{VLR}{Visitor Location Register}
\newacronym{hlr}{HLR}{Home Location Register}
\newacronym{auc}{AUC}{Authentication Center}
\newacronym{smsc}{SMSC}{The Short Message Service Center}
\newacronym{pstn}{PSTN}{Public Switched Telephone Network}
\newacronym{gmsc}{GMSC}{Gateway-MSC}
\newacronym{gprs}{GPRS}{General Packet Radio Service}
\newacronym{pdn}{PDN}{Packet Data Network}
\newacronym{pcu}{PCU}{Packet Control Unit}
\newacronym{sgsn}{SGSN}{Serving GPRS Support Node}
\newacronym{ggsn}{GGSN}{Gateway GPRS Support Node}
\newacronym{edge}{EDGE}{Edhanced Data Rates for GSM Evolution}
\newacronym{umts}{UMTS}{Universal Mobile Telecommunications System}
\newacronym{utran}{UTRAN}{UMTS Terrestrial Radio Network}
\newacronym{wcdma}{WCDMA}{Wideband Code Division Multiple Access}
\newacronym{rnc}{RNC}{Radio Network Controllers}
\newacronym{ue}{UE}{User Equipment}
\newacronym{hspa}{HSPA}{High Speed Packet Access}
\newacronym{lte}{LTE}{Long Term Evolution}
\newacronym{ofdm}{OFDM}{Orthogonal Frequency Division Multiplexing}
\newacronym{eutran}{E-UTRAN}{Evolved UTRAN}
\newacronym{enodeb}{eNodeB}{evolved Node B}
\newacronym{epc}{EPC}{Evolved Packet Core}
\newacronym{esim}{eSIM}{embedded Subscriber Identity Module}
\newacronym{mme}{MME}{Mobility Management Entity}
\newacronym{hss}{HSS}{Home Subscriber Service}
\newacronym{sgw}{SGW}{Serving Gateway}
\newacronym{pgw}{PGW}{Packet Data Network Gateway}
\newacronym{qos}{QoS}{Quality of Service}
\newacronym{qoe}{QoE}{Quality of Experience}
\newacronym{nf}{NF}{Network Function}
\newacronym{nfv}{NFV}{Network Function Virtualization}
\newacronym{amf}{AMF}{Access and Mobility Management Function}
\newacronym{upf}{UPF}{User Plane Function}
\newacronym{smf}{SMF}{Session Management Function}
\newacronym{gnodeb}{gNodeB}{next-generation Node B}
\newacronym{sim}{SIM}{Subscriber Identity Module}
\newacronym{grx}{GRX}{General Packet Radio Service (GPRS) Roaming eXchange}
\newacronym{ixp}{IXP}{Internet eXchange Point}
\newacronym{fno}{FNO}{Fixed Network Operator}
\newacronym{isp}{ISP}{Internet Service Provider}
\newacronym{asp}{ASP}{Application Service Provider}
\newacronym{sp}{SP}{Service Provider}
\newacronym[plural=PoPs, firstplural=Points of Presence (PoPs)]{pop}{PoP}{Point of Presence}
\newacronym{hr}{HR}{Home-Routed}
\newacronym{lbo}{LBO}{Local Breakout}
\newacronym{rbo}{RBO}{Regional Breakout}
\newacronym{ihbo}{IHBO}{IPX Hub Breakout}
\newacronym{imsi}{IMSI}{International Mobile Subscriber Identity}
\newacronym{isn}{ISN}{Initial Sequence Number}
\newacronym{urllc}{uRLLC}{ultra Reliable Low Latency Communication}
\newacronym{rtt}{RTT}{Round Trip Time}
\newacronym{ltem}{LTE-M}{LTE for Machines}
\newacronym{nbiot}{NB-IoT}{Narrowband IoT}
\newacronym{kpi}{KPI}{Key Performance Indicator}
\newacronym{m2m}{M2M}{Machine to Machine}
\newacronym{cdn}{CDN}{Content Delivery Network}
\newacronym{nr}{NR}{New Radio}
\newacronym{aws}{AWS}{Amazon Web Services}
\newacronym{mqtt}{MQTT}{Message Queuing Telemetry Transport}
\newacronym{mcc}{MCC}{Mobile Country Code}
\newacronym{mnc}{MNC}{Mobile Network Code}
\newacronym{mna}{MNA}{Mobile Network Aggregator}
\newacronym{mptcp}{MPTCP}{Multipath TCP}
\newacronym{gsma}{GSMA}{Global System for Mobile Communication (GSM) Association}
\newacronym{udp}{UDP}{User Datagram Protocol}
\newacronym{rat}{RAT}{Radio Access Technology}
\newacronym{ci}{CI}{Country Initial}
\newacronym{dns}{DNS}{Domain Name System}
\newacronym{sla}{SLA}{Service Level Agreement}
\newacronym{vpn}{VPN}{Virtual Private Network}
\newacronym{vlan}{VLAN}{Virtual Local Area Network}
\newacronym{vplmn}{VPMLN}{Visited Public Land Mobile Network}
\newacronym{hplmn}{HPMLN}{Home Public Land Mobile Network}
\newacronym{vmno}{v-MNO}{Visited Mobile Network Operator}
\newacronym{hmno}{h-MNO}{Home Mobile Network Operator}
\newacronym{bmno}{b-MNO}{Base Mobile Network Operator}
\newacronym{hrr}{HRR}{Home Routed Roaming}
\newacronym{cp}{CP}{Control Plane}
\newacronym{opex}{OPEX}{Operational Expenditures}
\newacronym{iaas}{IaaS}{Infrastructure-as-a-Service}
\newacronym{dlt}{DLT}{Distributed Ledger Technology}
\newacronym{dch}{DCH}{Data Clearing House}
\newacronym{cwnd}{cwnd}{congestion window}
\newacronym{pgpp}{PGPP}{Pretty Good Phone Privacy}

\begin{document}

\makeatletter
\def\endthebibliography{%
  \def\@noitemerr{\@latex@warning{Empty `thebibliography' environment}}%
  \endlist
}
\makeatother

\title{Challenges and Opportunities for Global Cellular Connectivity}

\settopmatter{authorsperrow=4}

\author{Viktoria Vomhoff}
\affiliation{
	\institution{University of Wuerzburg}
        \country{Germany}
}
\email{viktoria.vomhoff@uni-wuerzburg.de}

\author{Hyunseok Daniel Jang}
\affiliation{
	\institution{NYU Abu Dhabi}
        \country{UAE}
}
\email{daniel.jang@nyu.edu}

\author{Matteo Varvello}
\affiliation{
	\institution{Nokia Bell Labs}
        \country{USA}
}
\email{matteo.varvello@nokia.com}

\author{Stefan Gei\ss{}ler}
\affiliation{
	\institution{University of Wuerzburg}
        \country{Germany}
}
\email{stefan.geissler@uni-wuerzburg.de}

\author{Yasir Zaki}
\affiliation{
	\institution{NYU Abu Dhabi}
        \country{UAE}
}
\email{yasir.zaki@nyu.edu}

\author{Tobias Ho\ss{}feld}
\affiliation{
	\institution{University of Wuerzburg}
        \country{Germany}
}
\email{tobias.hossfeld@uni-wuerzburg.de} 

\author{Andra Lutu}
\affiliation{
	\institution{Telefonica Research}
        \country{Spain}
}
\email{andra.lutu@telefonica.com}

\begin{abstract}
Traditional cellular service was designed for global connectivity, but business and logistical constraints led to its fragmentation, with deployments limited to individual countries and regions. Initiatives like \glspl{mvno}, \glspl{mna}, and regulations like ``roam-like-at-home'' have partially restored global service potential, though often at high costs in terms of user bills, application performance, and traffic efficiency. This paper makes two key contributions: first, it surveys the global cellular ecosystem, analyzing the strengths and weaknesses of major players using data from prior research, proprietary datasets, and public sources. Second, it argues that the technology for seamless global service exists in \gls{lbo}, a roaming architecture which allows user traffic to be routed directly to the Internet through the visited network, bypassing the home network and/or third-party infrastructures. However, \gls{lbo} adoption is hindered by issues such as policy enforcement, billing, and \gls{qos} guarantees, rooted in a lack of trust between operators. The paper concludes by exploring technological advances that could enable \gls{lbo}, and pave the way for truly global cellular connectivity.


\end{abstract}



\settopmatter{printacmref=false, printfolios=false}
\maketitle

\section{Introduction}
\label{sec:intro}
\glsresetall
Traditional cellular service was designed for global connectivity, allowing mobile devices to function worldwide without the need of a separate service contract with a local operator in each visited country. This relies on a handful of telco providers that together with \glspl{mno} interconnect to offer cellular connectivity 
in a monolithic fashion, \ie a single entity provides all the components of the mobile communications service. 

As both the mobile market and the technology matured, \glspl{mvno}~\cite{schmitt2016study, xiao19mobisys, zarinni2014first} surfaced as a way to further exploit the infrastructure of incumbent \glspl{mno} to provide services without the need of deploying all the cellular network components. In recent years, \glspl{mna} -- such as Airalo~\cite{airalo}, 1Global~\cite{1Global} or Twilio~\cite{twilio} -- have emerged as a disruptive phenomenon in the telco sector to provide global mobile connectivity~\cite{alcala2022global, yuan18mobicom}. Their operational model upgrades the \gls{mvno} approach 
by leveraging the infrastructure of multiple base operators in different countries. 
To achieve this, \glspl{mna} overload the international roaming function, and benefit from the extensive global infrastructure that telcos have been shaping for decades.
Specifically, all \glspl{mna} gain access to (visited) \glspl{ran} globally via interconnection through roaming hubs~\cite{lutu2020first}.

Despite roaming, there is currently an intrinsic lack of trust between the \gls{hmno} and the \gls{vmno}, which ends up impacting communication performance. Specifically, the unwillingness of the \gls{hmno} to expose to a foreign operator charging information for their users makes \gls{hr} roaming the default roaming configuration. With \gls{hr}, all traffic is routed through the home country, regardless of the roaming user’s actual geolocation. This allows \glspl{mno} to control the charging function of their outbound roaming users, but it also translates into a non-negligible delay penalty and potential performance impairment~\cite{mandalari2018experience, fezeu2024roaming}.

However, despite these advancements, \glspl{mna} still face significant challenges in offering a native-like service to their roaming end-users.
In this paper, we highlight the need for a truly global network architecture for cellular connectivity, where the end-users can enjoy native-like 
performance.  We build on existing measurement studies~\cite{mandalari2018experience, mandalari2022measuring, alcala2022global, vomhoff2024shortcut, jang2024unravelingairaloecosystem} to map out the existing commercial implementations for global operators. 
We detail next the contributions we make in this work, as follows.

This paper presents 
a complete overview, at the time of writing, of \glspl{mna} models (Section~\ref{sec:mna_models}). Depending on their type (light, thick, full), \glspl{mna} achieve 
varying geo-spatial granularity of breakout locations.
Until now, we have seen commercial 
\glspl{mna} relying on the base operator's core networks (light \glspl{mna}) or fully/partially running their own (full, thick \glspl{mna}) to provision their eSIM profiles. 
With the advent of \gls{esim} technology -- which allows users to activate a cellular plan purely via software --  some \glspl{mna} now enable travelers to avoid inconvenient (buying local SIM cards in stores) and expensive (roaming bill shock) connectivity while abroad.  
The combination of \gls{mna} and \gls{esim} technology with recent advances in network virtualization~\cite{jain22L25GC, luo21cellbricks, larrea23coreKube} enables a new era in global mobile connectivity, characterized by varying degrees of operational complexity.

Despite creative in their setups, we explain that \glspl{mna} cannot currently offer a native-like service to their roaming end-users, for which the \gls{lbo} roaming architecture -- where traffic is directly offloaded at the \gls{vmno} -- is needed. We formulate 
the fundamental challenges that stand in the way of realizing \gls{lbo} within the cellular ecosystem (Section~\ref{sec:directions}). We highlight the corresponding research problems around zero-trust interaction, cross-operator security, reliable billing and ensuring consistent performance for the end-user.  Our contribution here aims to put emphasis on global connectivity research, specifically towards removing unrelenting limitations. 

We finally contribute a (partial) examination of potential avenues to achieve true global connectivity, and tackle (some of) the above-mentioned unrelenting challenges (see Section~\ref{sec:conclusion}). We discuss incremental solutions, such as amplifying the \gls{mna} model to leverage multiple cloud providers, or new architecture designs, such as attempting to solve the problem at the transport layer~\cite{luo21cellbricks}, or decoupling end-user identity from their provider~\cite{schmitt2021pretty}.
We note that these attempts ultimately introduce new actors into the ecosystem to facilitate or proxy the end-user's global access to cellular resources. Thus, we also explore solutions that enable \glspl{mno} to establish roaming partnerships and exchange value directly, eliminating the need for third parties~\cite{lutu2020dice}.

\section{Preliminaries}
\label{sec:background}


\vspace{0.05in}
\noindent \textbf{What is Global Cellular Connectivity?} Roaming is a fundamental feature of cellular networks that allows users to maintain connectivity outside their home network’s coverage area, ensuring seamless global service across various visited networks. 
Initially a niche service for occasional travelers, roaming is now the standard mode of operation for a growing amount of extremely heterogeneous devices that need to operate in different environments around the world, with varying performance requirements. 




\newcommand{\No}{\cellcolor{red!25}No\xspace}
\newcommand{\Yes}{\cellcolor{green!25}Yes\xspace}
\newcommand{\YesNo}{\cellcolor{orange!25}Yes*\xspace}
\newcommand{\NA}{\cellcolor[gray]{0.75}N/A\xspace}
\newcommand{\NoEmtpy}{\cellcolor{red!25}\xspace}
\newcommand{\YesEmtpy}{\cellcolor{green!25}\xspace}
\newcommand{\YesNoEmtpy}{\cellcolor{orange!25}\xspace}
\newcommand{\NAEmtpy}{\cellcolor[gray]{0.75}\xspace}

\begin{table*}[t]
\footnotesize
\setlength\tabcolsep{3pt}
\caption{\small MNA complexity: multiple actors (table columns) within the ecosystem come together to build the MNA's service. All aggregators depend on RAN providers, IPX Network and eSIM Platforms, and operate their own brand and sales. For Thick MNAs (emnify, Airalo), there are at least 6 different entities that coordinate to build the global connectivity service. The light MNA Google Fi only uses the cloud provider for breakout (orange cell) when the end-user activates the VPN service. The Full MNA Twilio operates their full core network over cloud infrastructure. In the "comments" we show the MNA roaming architecture and their target end-users.  }
\label{tab:mna_complexity}
\vspace{-2mm}
\begin{tabular}{c | c c c c c c  | c}
\toprule
    & \textbf{v-MNO} & \textbf{IPX Network} & 
    \textbf{Cloud Provider} & \textbf{b-MNO} &
    \textbf{eSIM Platform} & \textbf{MNA} & Comments \\ \hline \hline
\toprule
    Google Fi   & \Yes & \Yes   & \cellcolor{orange!25}VPN Tunnel Endpoint\xspace   & \cellcolor{red!25}Data Breakout & \Yes & \YesEmtpy Light MNA    & HR Roaming, Nomads~\cite{alcala2022global, yuan18mobicom} \\\hline 
    1Global     & \Yes & \Yes   & \NA                                               & \cellcolor{red!25}Data Breakout & \Yes & \YesEmtpy Light MNA    & HR Roaming, IoT + Nomads~\cite{alcala2022global} \\\hline 
    Airalo      & \Yes & \Yes   & \cellcolor{red!25}Data Breakout                   & \cellcolor{red!25}Data Breakout & \Yes & \YesEmtpy Thick MNA    & RBO + HR Roaming, Nomads~\cite{jang2024unravelingairaloecosystem} \\\hline   
    emnify      & \Yes & \Yes   & \cellcolor{red!25}Cloud Core, Data Breakout       & \Yes                            & \Yes & \YesEmtpy Thick MNA    & RBO, IoT~\cite{vomhoff2024shortcut} \\\hline  
    Twilio/KORE & \Yes & \Yes   & \cellcolor{red!25}Cloud Core, Data Breakout       & \NA                             & \Yes & \YesEmtpy Full MNA     & HR Roaming, IoT~\cite{alcala2022global} \\\hline    
\bottomrule
\end{tabular}
\end{table*}


This surge in demand for 
ubiquitous cellular connectivity comes both from the massive number of connected \gls{iot} devices as well as from people who are progressively switching (together with 
their devices) to a digital nomad lifestyle. Catering to digital nomads has led to the advent of global (virtual) operators that use roaming to provide seamless global connectivity, such as Airalo~\cite{airalo}, Holafly~\cite{holafly}, Yesim~\cite{yesim}, Nomad~\cite{nomad}, Onesim~\cite{onesim}, AloSIM~\cite{aloSIM}, Yoho Mobile~\cite{yoho}, BNESIM~\cite{bnesim}. At the same time, countless global \gls{iot} operators emerged
~\cite{lutu2020things}, such as emnify~\cite{emnify}, KORE~\cite{KORE} or 1Global~\cite{1Global}.  
Unlike human-generated traffic, \gls{iot} traffic involves smaller, more frequent data exchanges which often require multiple signaling messages for minimal data transmission~\cite{geissler2024untangling}. Efficient network planning and management for transporting signaling traffic are essential to maintaining performance in future next-generation networks, particularly given the rapid proliferation of \gls{iot} devices. 


\vspace{0.05in}
\noindent \textbf{Architectures, Agreements and Billing:} Seamless global connectivity relies on cooperation among at least three independent stakeholders: visited network (providing the radio access), home network (managing the user's identity), and interconnection provider. This cooperation is supported by standardized interfaces and protocols but also requires significant administrative work via roaming agreements. The home network, responsible for providing the SIM identity and connectivity, serves a user's device within its own geographical region. When the user travels outside this area, their device connects to the visited network’s radio access via roaming. 

\begin{figure}[tb]
	\centerline{\includegraphics[width=0.99\columnwidth]{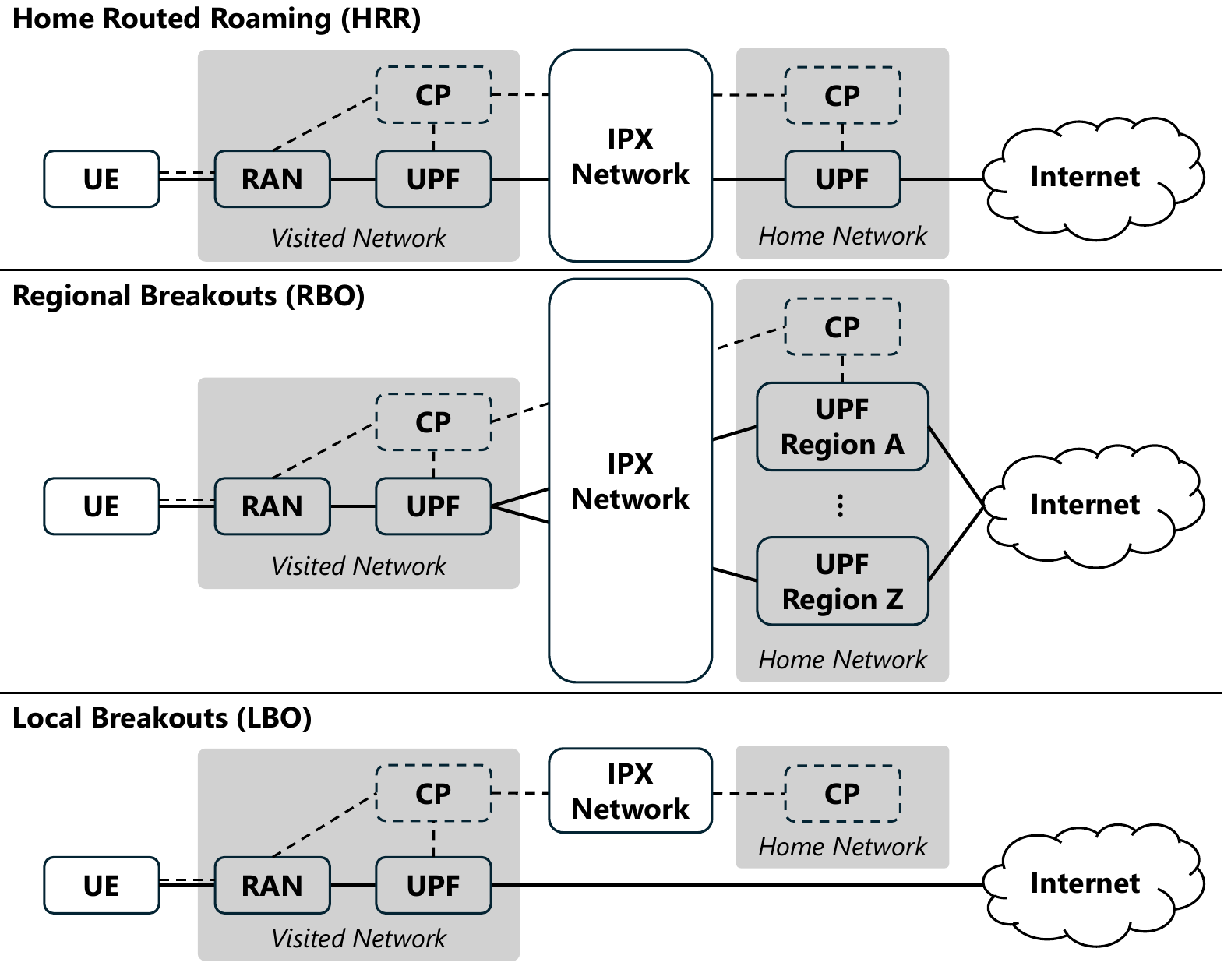}}
	\Description{}
        \vspace{-5pt}
        \caption{Simplified 5G roaming architecture. Dashed lines: control plane traffic; solid lines: user plane traffic.}
	\label{fig:roaming}
\end{figure}

Communication between home and visited networks typically occurs via the \gls{ipx} network~\cite{lutu2021insights}, traversing at most two separate carriers~\cite{vomhoffshortcut,lutu2020things}. Whether signaling traffic, user plane traffic, or both are transmitted between visited and home networks depends on the specific roaming architecture in place. Figure~\ref{fig:roaming} illustrates three 5G roaming architectures: \gls{hrr}, where all user traffic is routed through the home network, ensuring maximum control but at the cost of higher latency; \gls{rbo}, which balances control and efficiency by offloading traffic at regional points closer to the user; and \gls{lbo}, where traffic is directly offloaded at the visited network, minimizing latency but reducing the home network's control over service quality and security.
To ensure smooth international mobile communications, the so-called clearing houses play a crucial role by ensuring correct processing, validation, billing and accounting of mobile usage data across the different networks~\cite{macia2009fraud}.

\begin{figure}[!t]
    \includegraphics[width=\columnwidth]{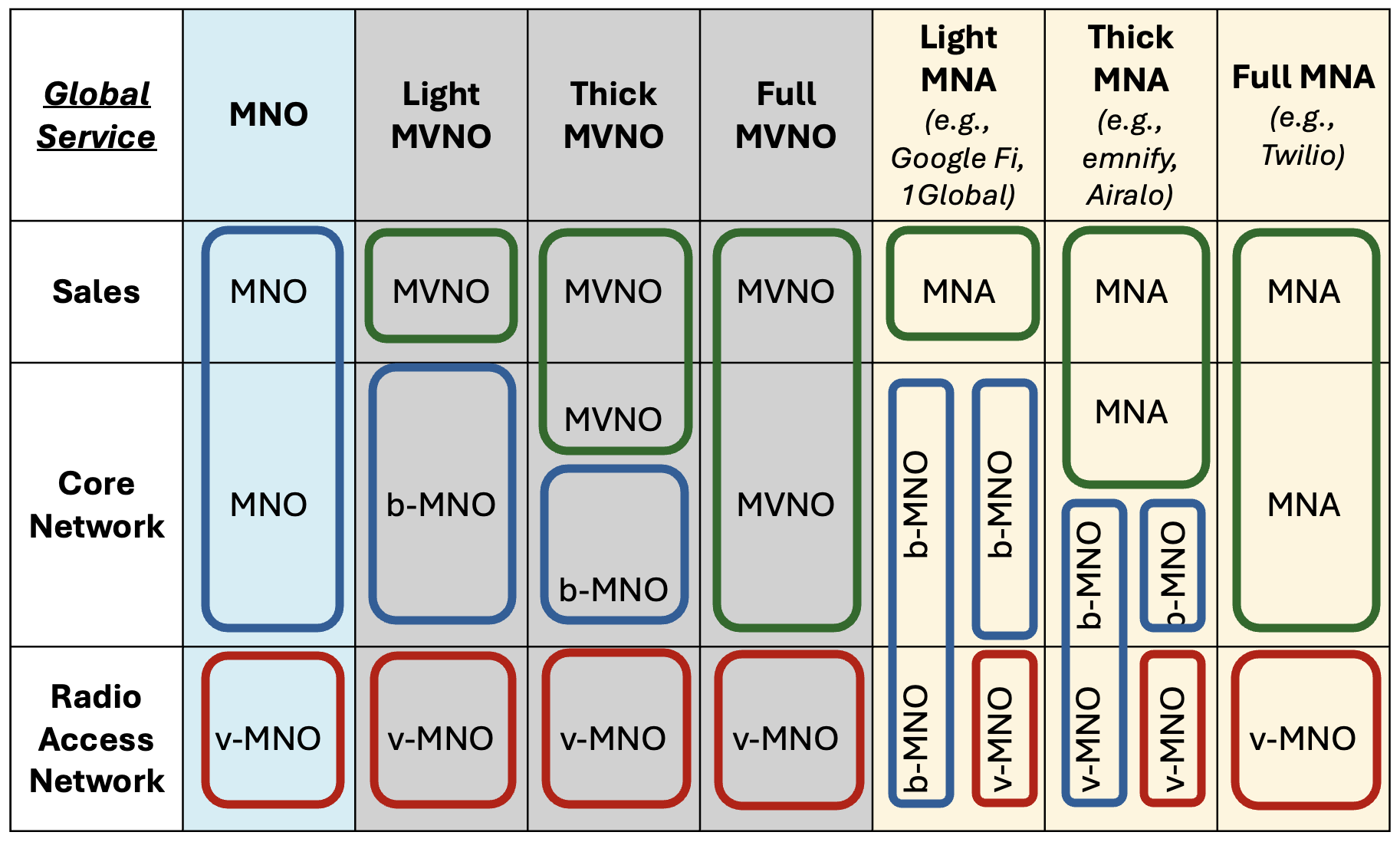}
    \caption{\small Network operating models: MVNOs lacks ownership of radio spectrum resources; light/thick MVNOs rely on a single b-MNO, and the latter's roaming agreements for global service, while full MVNOs operate their own core network, and give global service through their own roaming agreements.
    MNAs run a limited part of the network (the \textit{light} -- only sales $<<$  the \textit{thick} -- limited core function $<<$ the \textit{full} -- all the core), and provide global service by exploiting the roaming agreements of several b-MNOs. }
    \label{fig:mna_models}
\end{figure}

\vspace{0.05in}
\noindent \textbf{Mobile Virtual Network Enablers and Roaming:}
The conjunction of 
novel network softwarization and virtualization technologies -- specifically eSIM technology and control and data plane separation in 5G networks -- paired with the extent of cloud infrastructure deployment 
led to the realization of numerous global operators (be it for IoT verticals or digital nomads) -- which we generically denote as \glspl{mna}.
Figure~\ref{fig:mna_models} represents the different flavors of virtual operators, either global or restricted to a single geography.




%

\glspl{mna} evolve from the \glspl{mvno} operator model~\cite{zarinni14imc}, which allows new operators to rent infrastructure from established \glspl{mno} to lower entry barriers. Unlike \glspl{mno}, \glspl{mvno}  provide mobile services without owning or operating a full cellular network, specifically lacking radio spectrum ownership. To operate, an \gls{mvno} 
must establish commercial agreements to 
use the network of a base \gls{mno}. 
Several \gls{mvno} types exist, depending on the extent of the technological layer 
added over the base \gls{mno}'s network resources~\cite{schmitt16pam, xiao19mobisys}. A 
``thin'' \gls{mvno} relies on a base \gls{mno}'s \gls{ran} and core network, while managing its own customer support, marketing, and pricing. A ``thick'' \gls{mvno} partially operates its own core network infrastructure for greater control but still relies on a base \gls{mno} for some functions, focusing on brand differentiation and additional services.  A ``full'' \gls{mvno} operates its own core network, relying on a base \gls{mno} only for access to radio resources.

The underlying infrastructure that supports \glspl{mvno} is the \gls{ipx} network, which interconnects virtually all \glspl{mno} world-wide~\cite{lutu2020first} (even the ones within the same economy).
In other words, the very same interconnections within the cellular ecosystem that enable (national/international) roaming are the ones that also supports the realization of \glspl{mvno}. 
\gls{ipx} providers act as mobile virtual network \textit{enablers}, and the activation of a full \gls{mvno} is similar to that of an international roaming agreement for \gls{hrr}. 



\glspl{mna} rely on the core networks of base operators (\textit{Light \glspl{mna}})~\cite{GoogleFi, 1Global} or run their own (\textit{Full \glspl{mna}})~\cite{alcala2022global, yuan18mobicom} to provision their \gls{esim} profiles. Both models gain access to (visited) radio access networks globally via interconnection through roaming hubs~\cite{lutu2020first}. \textit{Thick MNAs} push this model further, and decouple the internet gateway location from both the base \glspl{mno}' and the visited operators' infrastructure. For example, Airalo~\cite{airalo} is a popular thick \gls{mna}~\cite{jang2024unravelingairaloecosystem}, which has gained more than 5 million customers
since its inception in 2019. Emnify~\cite{emnify} leverage a similar operational model focused on IoT-specific global connectivity: where they use seven different cloud locations as roaming breakout points, all from the same provider.

\section{Reality Check on Global Cellular Connectivity}
\label{sec:mna_models}

We dissect here the commercial implementations of \textit{global} 
cellular connectivity
which we identified at the time of writing. 



\subsection{Evolving From \gls{hr} Roaming}

Global cellular service relies on international roaming, with \gls{hr} being the preferred setup, even for 5G networks~\cite{fezeu2024roaming}. 
\glspl{mna} were designed to 
reduce the inefficiencies associated with \gls{hr}, and improve performance by dynamically switching between \glspl{bmno} from different countries, thus reducing the delay penalty of \gls{hr}.

\glspl{mna} are still built on the same trust model as international roaming, while 
allowing for service delivery 
almost 
worldwide.
By aggregating network resources and functions provided by different actors (e.g., \gls{ran} resources from different \glspl{vmno}, data breakout from the \glspl{bmno}), \glspl{mna} implicitly fragment the end-to-end service across different network domains. 
Switching the \gls{bmno} used by an \gls{mna} results in a different location for data breakout to the Internet, while still relying on the \gls{hr} roaming architecture, where the ``home'' is the respective \gls{bmno}. The \gls{bmno} choice 
varies based on factors like policy, coverage, or performance.

\begin{figure*}
    \centering
    \includegraphics[width=0.9\linewidth]{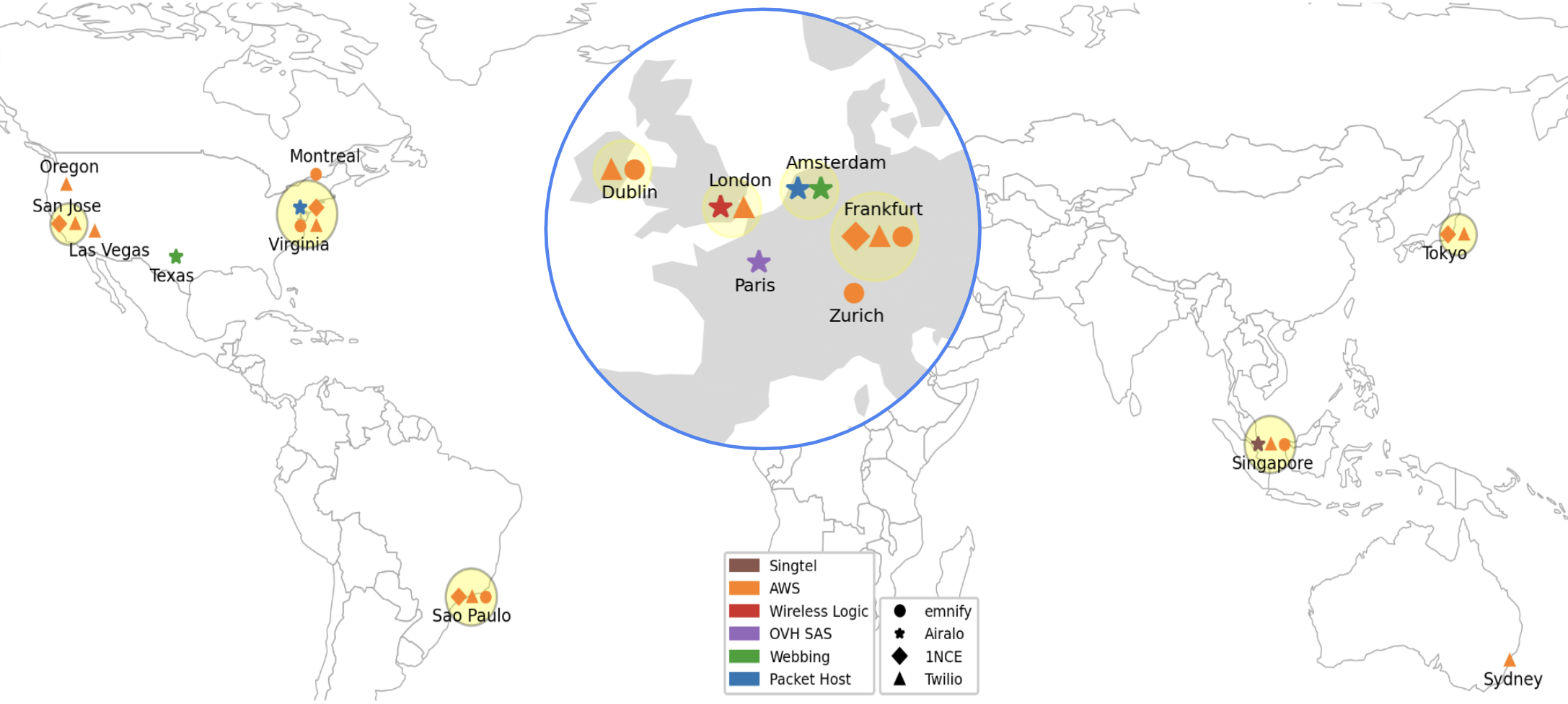}
    \caption{\small Map showing the geolocation of public internet gateways for various commercial eSIM providers—1NCE, Airalo, emnify, and Twilio—represented by different symbols. Colors denote the network providers hosting these gateway functions.}
    \label{fig:gw_map}
\end{figure*}
\subsection{Mapping MNA Breakout}
Previous work conducted active measurements to map breakout locations given variable measurement vantage points and \glspl{mna}~\cite{mandalari2018experience, mandalari2022measuring, alcala2022global, vomhoff2024shortcut}. The findings of these studies unveil how \glspl{mna} are gradually exploiting the idea of swapping \gls{bmno} (and implicitly their corresponding breakout locations) or -- in the case of thick \glspl{mna} -- deploying internet gateways independently from the \gls{bmno} in breakout locations closer to the end-user, all in an effort to reduce the implications of the \gls{hr} roaming setup~\cite{mandalari2022measuring, fezeu2024roaming}.
Table~\ref{tab:mna_complexity} gives an overview of the way in which specific MNAs (namely, Google Fi, 1Global (previously known as Truphone), Airalo, emnify and Twilio) build their service. 

Depending on their setup (see classification in Table~\ref{tab:mna_complexity} according to the taxonomy from Section~\ref{sec:background}), \glspl{mna} achieve different levels of geo-spatial granularity in terms of the breakout locations.
Figure~\ref{fig:gw_map} illustrates the geographic distribution of locations hosting internet gateway functions 
utilized by some commercial \glspl{mna}. These gateways deploy across different network providers, including cloud service providers (AWS, Packet Host, OVH SAS), a telecom carrier (Singtel), and IoT platforms (Webbing, Wireless Logic). We use public documentation to collect this data for 1NCE, emnify, and Twilio, which specifies the AWS regions hosting their gateways. For Airalo, we use the results of a measurement study with volunteers using Airalo eSIMs across different countries to infer (some of) the network host and geolocation data~\cite{jang2024unravelingairaloecosystem}.

\vspace{0.05in}
\noindent
\textbf{Continent-level breakout (e.g., Google Fi)}: 
Google Fi is a \textit{light MNA} that relies on T-Mobile in the US, and Three UK in Europe~\cite{alcala2022global}.
Fi targets users in the US, and also offers global roaming coverage. 
For users traveling from the US to Spain, Fi swapped the base \gls{mno} from T-Mobile to Three UK. 
The use of \gls{rbo} in Europe on top of Three's network helps Fi to reduce significantly (i.e., from $\approx$200ms to $\approx$75ms) the delay their users experience in Europe, whenever the VPN service is not active. This service allows users to maintain the same online experience abroad as they have at home, including access to geo-restricted content. 

\vspace{0.05in}
\noindent
\textbf{Country-level breakout (e.g., 1Global)}: 1Global (previously known as Truphone) is a \textit{light MNA} that aggregates separate individual MVNO agreements from the nine economies where they register. 
1Global was, in fact, the closest -- according to measurement campaign in~\cite{alcala2022global} -- to deliver native-like mobile connectivity to global users. 
This is achieved by mapping the breakout point to the closest \gls{mvno} partnerships to the aimed service location. 
Leveraging their mature setup with different MVNO partnerships in over nine economies, 1Global delivers the closest performance to the one provided by a local \gls{mno} in any of those countries. 

\vspace{0.05in}
\noindent\textbf{Sponsored Roaming and Cloud Breakout (\eg emnify, Twilio
, Airalo):} emnify is a cloud-based platform for global cellular IoT connectivity, allowing businesses to manage IoT devices across networks without multiple SIM cards or \gls{mno} contracts. It operates seven breakout regions worldwide -- hosted on AWS -- where data exits the cellular network to the internet or private networks, reducing latency and enhancing data transmission speed~\cite{vomhoff2024shortcut}.

Airalo is a prominent example of a \textit{thick MNA}, leveraging the deployment of internet breakout gateways in multiple third-party (cloud) network infrastructures that are decoupled from both the base and the visited \glspl{mno}. In theory, this mode of operation allows for the benefits of \gls{rbo}, particularly the dynamic traffic routing and prioritization of \glspl{upf} closer to the user. In practice, Airalo eSIMs are configured to statically route traffic through specific internet gateways, often resulting in geographically suboptimal data paths. For example, its eSIMs for Azerbaijan, Finland, Moldova, and Kenya use Telecom Italia as \gls{bmno}~\cite{jang2024unravelingairaloecosystem}, but their roaming traffic consistently relies on \gls{rbo} via \glspl{upf} in London, operated by Wireless Logic. Airalo's performance measurement across 21 countries found that while \gls{rbo} eSIMs offered better latency compared to \gls{hr} counterparts, they showed minimal improvement in bandwidth~\cite{jang2024unravelingairaloecosystem}. This highlights the challenge of \gls{rbo} operations due to complex interplay of agreements among \glspl{mno}, \gls{ipx} providers, and network infrastructure companies.

Finally, Twilio is a \textit{full MNA} that operate their own cloud-ified core network~\cite{alcala2022global}, and their own \gls{mcc}-\gls{mnc}. It uses \gls{hr} roaming to their own core locations hosted in cloud infrastructure. 


\addtolength{\tabcolsep}{-3pt} 
\begin{table*}[ht]\centering
	\caption{Challenges and opportunities of Local Breakouts (LBO) for international mobile roaming.}
	\label{tab:summary}
	\vspace{-3mm}
	\begin{tabulary}{\textwidth}{@{}lLL@{}}
            \toprule
            \textbf{Aspect} & \textbf{Challenges} & \textbf{Opportunities} \\
            \midrule
            \multirow{2}{1,5cm}{Trust \& Billing} & The lack of trust between entities regarding security, authorization, or billing hinders efficient cooperation and optimal service performance. & Develop standardized zero-trust frameworks for secure inter-operator interactions in 6G networks, to enhance global cellular connectivity, prevent roaming billing disputes, and support dynamic billing models. \\
            \multicolumn{3}{c}{\textbf{RQ: How can trust between parties be ensured on a technical level?}} \\\hline
            %
            %
             \multirow{2}{1,7cm}{Security \& Service Monitoring}: & Maintaining consistent security and privacy policies across independent networks and operators. & Potential to implement specialized and localized security measures (enabled via. e.g, global standardized APIs, standardized network data representation), tailored to specific regional requirements. \\
            \multicolumn{3}{c}{\textbf{RQ: How can specialized security and service monitoring mechanisms be deployed across operator boundaries?}} \\\hline
            Regulatory & Complying with different regulatory requirements in various regions is complex and resource-intensive. & Enhanced transparency and credibility in carbon footprint reporting, improving stakeholder trust. \\
             & \multicolumn{2}{c}{\textbf{RQ: How can compliance with diverse regulatory environments be managed efficiently?}} \\\hline
            \multirow{2}{1,7cm}{Quality of Service (QoS)} & Potential variability in service quality across different visited networks. & Improved QoS for local content and services due to reduced latency and localized traffic handling. \\
             & \multicolumn{2}{c}{\textbf{RQ: How can QoS be improved to ensure consistent service quality across visited networks?}} \\\hline
            %
            %
            %
            \multirow{2}{1,7cm}{Application Layer} & Impact on services due to geolocation and distance from home country and content. & Utilizing advanced techniques to handle data locally, thereby reducing latency and improving service performance. \\
             & \multicolumn{2}{c}{\textbf{RQ: How can localized data handling mitigate the impact of geolocation and distance on services?}} \\\hline
            %
	\end{tabulary}
\end{table*}
\addtolength{\tabcolsep}{+3pt} 

\section{LBO for Global Connectivity}
\label{sec:directions}

The previous section highlighted inefficiencies in current \glspl{mna}' solutions for global cellular connectivity. We identify \gls{lbo} as a promising technological advancement to address performance issues in \gls{rbo} or \gls{hr} architectures. This section explores the migration from \gls{hr} to \gls{lbo} roaming, outlining the associated opportunities and challenges within the cellular ecosystem. Table~\ref{tab:summary} provides a concise overview, including research questions for each aspect.

\vspace{0.05in}
\noindent\textbf{Trust \& Billing:} 
Although \glspl{lbo} are technically feasible, their adoption is hindered by a lack of trust within the mobile ecosystem. Third-party networks may not follow the same security standards, risking data breaches and unauthorized access~\cite{liyanage2016opportunities}. Trust is essential for accurate billing, financial settlements, and fraud prevention, requiring precise data tracking and real-time reconciliation between home and visited networks~\cite{macia2009fraud}.

Implementing zero-trust frameworks, distributed ledgers for transparent transactions~\cite{refaey2019blockchain, mafakheri2021smart, elmadani2022blockchain}, advanced cryptography~\cite{sadhukhan2024development}, and international regulatory harmonization~\cite{prasad20235g} can address these challenges by creating standardized trust and security frameworks. Regulatory support can further enhance trust and compliance.

\vspace{0.05in}
\noindent
\textbf{Security \& Service Monitoring:} 
Trust issues between mobile entities raise concerns about traffic handling and privacy, particularly in \gls{lbo}, where user traffic is controlled by visited networks with varying security standards~\cite{xie2014privacy, bikos2012lte}. Strong authentication and authorization are crucial~\cite{mahyoub2024security, khan2019survey, chakraborty2023framework}. 

Shifting control to the visited network offers opportunities for region-specific security optimization~\cite{liyanage2016opportunities}, enabling better protection against local threats and easier compliance with regional regulations. However, the key challenge lies in developing methods and architectures that allow home network functions to be effectively carried out within the visited network.

\vspace{0.05in}
\noindent\textbf{Regulatory Aspects:} Regulatory compliance, especially with data sovereignty laws like GDPR~\cite{voigt2017eu}, complicates \gls{lbo} deployments, as these laws vary by country. Ensuring local breakouts meet both home and visited countries' legal requirements, including lawful interception, is challenging. Previously monitored by the home network, accessing the same data in \gls{lbo} deployments is not straightforward~\cite{dhar2024blockchain}, requiring clear data ownership in roaming agreements.

Another key regulatory concern is the reporting of CO2 emissions within the corporate carbon footprint~\cite{klaassen2021harmonizing}. Companies must report the following CO2 emissions: Scope 1 (direct emissions from owned sources), Scope 2 (emissions from purchased energy), and Scope 3 (all other indirect emissions in the value chain). Therefore, the classification of emissions from roaming customers needs to be clarified, as they could be Scope 1 for the visited network or Scope 2 or 3 for the home operator.

\vspace{0.05in}
\noindent \textbf{Quality of Service:}  Varying \gls{qos} standards among \glspl{mno} complicate establishing \glspl{sla} that ensure consistent \gls{qos} for users~\cite{vomhoff2024shortcut}. Although\glspl{lbo} may offer higher \gls{qos} than \gls{hrr} (see Section~\ref{sec:background}), the home operator loses \gls{qos} control, with the visited network only providing \gls{qos} level the user pays for. Further, \gls{qos} gains may be less evident for services hosted near their home network.

\glspl{lbo} also complicate the support of proprietary features such as VPNs or real-time data tracking, typically managed by the home network~\cite{emnifyIntegrationGuides}. To address this, visited networks could offer \gls{iaas} components, allowing home operators to deploy custom features, like region-specific content filtering or localized firewall services. However, this approach risks fragmentation and inconsistencies, as each visited network may implement such features differently.

\vspace{0.05in}
\noindent 
\textbf{Application Layer:} Transitioning to \glspl{lbo} means assigning roaming devices public IP addresses from the visited network, affecting services like IP geolocation and content delivery. This can impact service performance, user experience, and application functionality, with users potentially receiving localized content or experiencing higher latency depending on where a service is hosted. While local breakouts improve efficiency for geo-distributed services (e.g. on CDNs), they might have little to no, or even negatively impact, applications hosted near the home network. However, localized data handling can optimize service delivery based on regional needs.

\section{Towards True Global Service}
\label{sec:conclusion}

Traditionally, 
cellular service relies on the \gls{imsi}, a permanent and globally-unique identifier that is stored on a \gls{sim}
card for both billing and authentication functionality as well as mobility and connectivity.
True global aggregation of cellular resources should allow end-users (and their \gls{imsi}) to access radio networks on-demand from any available operator world-wide -- regardless of size or trust level -- and do so in real-time.
Attempting to achieve this in a setup where the end-user identity is akin to the (sole) connectivity provider issuing the \gls{imsi} is restrictive.

We next explore both incremental and clean-slate approaches to achieving true global service, including innovative cellular architectures that eliminate existing barriers to multi-network access for users worldwide.

\vspace{0.05in}
\noindent
\textbf{Amplifying the Thick MNA Model:} 
With the materialization of \glspl{mna} 
such as Google Fi, 1Global, emnify, Airalo or Twillio, the cellular ecosystem evolved in terms of complexity, with data paths that were once confined to a single operator realm now traversing multiple domains, and relying on resources from different entities (including the \gls{ran} provider, the \gls{bmno}, the gateway provider, or \gls{ipx} provider).
Although creative, these operators still built on the very same trust model of the cellular ecosystem, and thus suffer from the limitations of combining international roaming and virtual operator design (see Section~\ref{sec:directions}). 
One avenue to explore is evolving the \textit{thick MNA} model to dynamically deploy Internet gateways over the underlying infrastructure of multiple providers.
Figure~\ref{fig:gw_map} highlights an existing concentration of internet breakout locations in specific cities, such as Frankfurt, Sao Paulo, Singapore, and Virginia, indicating these cities as major hubs for global roaming connectivity. 
This incremental solution could enhance the geospatial granularity of breakout locations, potentially achieving \gls{lbo}-like (native) performance (see Section~\ref{sec:directions}).

\vspace{0.05in}
\noindent
\textbf{Cellular Aggregation at the Transport Layer:} One promising approach to achieving global service is aggregating cellular networks at the transport layer, using multiple networks simultaneously to provide data service and mask the interruptions of any single network. 
Early research trying to apply \gls{mptcp} directly onto two cellular networks in the high speed mobility case points out that this vanilla attempt achieves little enhancement, or even occasionally performed worse compared to a single good path, due to imbalanced \gls{cwnd} distribution by coupled congestion control algorithm and an exaggerated out-of-order problem from frequent handovers~\cite{li2018measurement}.

Combining architecture design with aggregation at the transport layer and \gls{dlt}~\cite{romero2018distributed}, CellBricks~\cite{luo21cellbricks} proposed to move support for mobility from the network to the user device, so that a user can experience seamless mobility, even if they frequently switch between mobile providers.  However, these solutions require several changes, including modifications to cellular core software functions, updates to \gls{ue} firmware, and configuration changes to enable \gls{mptcp} in the network software stack on both clients and servers. 

\vspace{0.05in}
\noindent
\textbf{Decoupling End-user Identity From Their Provider:} Another potential avenue to enable global access to cellular networks is decoupling the identity of the end-user from the cellular infrastructure provider. 
\gls{pgpp}~\cite{schmitt2022decoupling, schmitt2021pretty} decouples billing and authentication from the cellular core, altering it to use an over-the-top oblivious authentication protocol to an external server, also supported through \gls{dlt}. This can be operated by a second organization, while leaving mobility and connectivity functions in the core as they were. 


\vspace{0.05in}
\noindent
\textbf{\gls{mno} Consortium:} The above-mentioned approaches combine the innovation potential of \gls{dlt} with architectural changes that enable global cellular service. 
However, they all propose the introduction of new actors within the ecosystem, which would support or act as proxies for the end-user's true global access to cellular resources. 
We explore here solutions that allow \glspl{mno} to engage in a roaming partnership and to exchange value easily, without the need of a third party to act as a trusted intermediary with the role to verify the interaction between the roaming partners.

The solution of cooperation among the \glspl{mno} through dynamic interconnection in the cellular ecosystem (DICE) simplifies the current roaming architecture, and allows for decentralized authentication of end-users looking to access resources globally~\cite{lutu2020dice}. 
The shift in the business logic that DICE brings through the use of cryptocurrencies allows for zero-trust global charging, without leaving the doubt that fraud (e.g., tampering with roaming records) might occur, thus promoting the local breakout roaming configuration.
DICE makes a two-fold contribution to disrupt the cellular ecosystem: (i) enables \glspl{mno} to have a direct dynamic cooperation in a private and secure setup, establish roaming relationships, and perform data and financial clearing in almost real time, and (ii) enables the end-user to have control over her mobile connection and connect to a network in a visited country as a native user, receiving optimal service performance. 
The \gls{dlt} solution lowers uncertainty about the identity of the different entities involved in the roaming transaction (i.e, the roaming user, the home network and the visited network offering roaming services to the roaming user), and allows the exchange of value without trust between the entities.


\balance

\bibliographystyle{ACM-Reference-Format} 
\bibliography{literature}	


\begin{thebibliography}{54}


\ifx \showCODEN    \undefined \def \showCODEN     #1{\unskip}     \fi
\ifx \showDOI      \undefined \def \showDOI       #1{#1}\fi
\ifx \showISBNx    \undefined \def \showISBNx     #1{\unskip}     \fi
\ifx \showISBNxiii \undefined \def \showISBNxiii  #1{\unskip}     \fi
\ifx \showISSN     \undefined \def \showISSN      #1{\unskip}     \fi
\ifx \showLCCN     \undefined \def \showLCCN      #1{\unskip}     \fi
\ifx \shownote     \undefined \def \shownote      #1{#1}          \fi
\ifx \showarticletitle \undefined \def \showarticletitle #1{#1}   \fi
\ifx \showURL      \undefined \def \showURL       {\relax}        \fi
\providecommand\bibfield[2]{#2}
\providecommand\bibinfo[2]{#2}
\providecommand\natexlab[1]{#1}
\providecommand\showeprint[2][]{arXiv:#2}

\bibitem[1Global({[n.\,d.]})]%
        {1Global}
\bibfield{author}{\bibinfo{person}{1Global}.} \bibinfo{year}{[n.\,d.]}\natexlab{}.
\newblock \bibinfo{title}{Global telecom solutions to scale your Business}.
\newblock \bibinfo{howpublished}{{ https://www.1global.com/}}.
\newblock


\bibitem[Airalo({[n.\,d.]})]%
        {airalo}
\bibfield{author}{\bibinfo{person}{Airalo}.} \bibinfo{year}{[n.\,d.]}\natexlab{}.
\newblock \bibinfo{title}{{Buy eSIMs for international travel}}.
\newblock \bibinfo{howpublished}{https://www.airalo.com/}.
\newblock


\bibitem[Alcal{\'a}-Mar{\'\i}n et~al\mbox{.}(2022)]%
        {alcala2022global}
\bibfield{author}{\bibinfo{person}{Sergi Alcal{\'a}-Mar{\'\i}n}, \bibinfo{person}{Aravindh Raman}, \bibinfo{person}{Weili Wu}, \bibinfo{person}{Andra Lutu}, \bibinfo{person}{Marcelo Bagnulo}, \bibinfo{person}{Ozgu Alay}, {and} \bibinfo{person}{Fabi{\'a}n Bustamante}.} \bibinfo{year}{2022}\natexlab{}.
\newblock \showarticletitle{Global mobile network aggregators: taxonomy, roaming performance and optimization}. In \bibinfo{booktitle}{\emph{Proceedings of the 20th Annual International Conference on Mobile Systems, Applications and Services}}. \bibinfo{pages}{183--195}.
\newblock


\bibitem[AloSIM({[n.\,d.]})]%
        {aloSIM}
\bibfield{author}{\bibinfo{person}{AloSIM}.} \bibinfo{year}{[n.\,d.]}\natexlab{}.
\newblock \bibinfo{title}{{Up to 85\% off prepaid eSIM data in 175+ countries}}.
\newblock \bibinfo{howpublished}{https://alosim.com/}.
\newblock


\bibitem[Bikos and Sklavos(2012)]%
        {bikos2012lte}
\bibfield{author}{\bibinfo{person}{Anastasios~N Bikos} {and} \bibinfo{person}{Nicolas Sklavos}.} \bibinfo{year}{2012}\natexlab{}.
\newblock \showarticletitle{LTE/SAE security issues on 4G wireless networks}.
\newblock \bibinfo{journal}{\emph{IEEE Security \& Privacy}} \bibinfo{volume}{11}, \bibinfo{number}{2} (\bibinfo{year}{2012}), \bibinfo{pages}{55--62}.
\newblock


\bibitem[Chakraborty et~al\mbox{.}(2023)]%
        {chakraborty2023framework}
\bibfield{author}{\bibinfo{person}{Pousali Chakraborty}, \bibinfo{person}{Marius Corici}, \bibinfo{person}{Hemant Zope}, \bibinfo{person}{Carlos Barjau}, \bibinfo{person}{Muhammad~Faheem Awan}, \bibinfo{person}{Josep Ribes}, \bibinfo{person}{Aaron~Montilla Vicent}, \bibinfo{person}{David Gomez-Barquero}, {and} \bibinfo{person}{Thomas Magedanz}.} \bibinfo{year}{2023}\natexlab{}.
\newblock \showarticletitle{A Framework for Roaming between 5G Non-Public-Networks (NPNs)}. In \bibinfo{booktitle}{\emph{2023 IEEE Conference on Standards for Communications and Networking (CSCN)}}. IEEE, \bibinfo{pages}{247--253}.
\newblock


\bibitem[Dhar~Dwivedi et~al\mbox{.}(2024)]%
        {dhar2024blockchain}
\bibfield{author}{\bibinfo{person}{Ashutosh Dhar~Dwivedi}, \bibinfo{person}{Rajani Singh}, \bibinfo{person}{Keshav Kaushik}, \bibinfo{person}{Raghava Rao~Mukkamala}, {and} \bibinfo{person}{Waleed~S Alnumay}.} \bibinfo{year}{2024}\natexlab{}.
\newblock \showarticletitle{Blockchain and artificial intelligence for 5G-enabled Internet of Things: Challenges, opportunities, and solutions}.
\newblock \bibinfo{journal}{\emph{Transactions on Emerging Telecommunications Technologies}} \bibinfo{volume}{35}, \bibinfo{number}{4} (\bibinfo{year}{2024}), \bibinfo{pages}{e4329}.
\newblock


\bibitem[Elmadani et~al\mbox{.}(2022)]%
        {elmadani2022blockchain}
\bibfield{author}{\bibinfo{person}{Safwan Elmadani}, \bibinfo{person}{Salim Hariri}, {and} \bibinfo{person}{Sicong Shao}.} \bibinfo{year}{2022}\natexlab{}.
\newblock \showarticletitle{Blockchain Based Methodology for Zero Trust Modeling and Quantification for 5G Networks}. In \bibinfo{booktitle}{\emph{2022 IEEE/ACS 19th International Conference on Computer Systems and Applications (AICCSA)}}. IEEE, \bibinfo{pages}{1--9}.
\newblock


\bibitem[Emnify({[n.\,d.]})]%
        {emnify}
\bibfield{author}{\bibinfo{person}{Emnify}.} \bibinfo{year}{[n.\,d.]}\natexlab{}.
\newblock \bibinfo{title}{{IoT Connectivity As Unique As Your Business}}.
\newblock \bibinfo{howpublished}{https://www.emnify.com/}.
\newblock


\bibitem[emnify(2023)]%
        {emnifyIntegrationGuides}
\bibfield{author}{\bibinfo{person}{emnify}.} \bibinfo{year}{2023}\natexlab{}.
\newblock \bibinfo{title}{{I}ntegration guides | emnify {D}ocumentation --- docs.emnify.com}.
\newblock \bibinfo{howpublished}{\url{https://docs.emnify.com/integration-guides}}.
\newblock
\newblock
\shownote{[Accessed 31-07-2024]}.


\bibitem[Ever({[n.\,d.]})]%
        {bnesim}
\bibfield{author}{\bibinfo{person}{BNE: Best~Network Ever}.} \bibinfo{year}{[n.\,d.]}\natexlab{}.
\newblock \bibinfo{title}{{Get your eSIM Stay connected wherever you are}}.
\newblock \bibinfo{howpublished}{https://my.bnesim.com/}.
\newblock


\bibitem[Fezeu et~al\mbox{.}(2024)]%
        {fezeu2024roaming}
\bibfield{author}{\bibinfo{person}{Rostand~AK Fezeu}, \bibinfo{person}{Claudio Fiandrino}, \bibinfo{person}{Eman Ramadan}, \bibinfo{person}{Jason Carpenter}, \bibinfo{person}{Daqing Chen}, \bibinfo{person}{Yiling Tan}, \bibinfo{person}{Feng Qian}, \bibinfo{person}{Joerg Widmer}, \bibinfo{person}{Zhi-Li Zhang}, {et~al\mbox{.}}} \bibinfo{year}{2024}\natexlab{}.
\newblock \showarticletitle{Roaming across the European Union in the 5G Era: Performance, Challenges, and Opportunities}. In \bibinfo{booktitle}{\emph{IEEE International Conference on Computer Communications}}.
\newblock


\bibitem[Gei{\ss}ler et~al\mbox{.}(2024)]%
        {geissler2024untangling}
\bibfield{author}{\bibinfo{person}{Stefan Gei{\ss}ler}, \bibinfo{person}{Andra Lutu}, \bibinfo{person}{Florian Wamser}, \bibinfo{person}{Thomas Favale}, \bibinfo{person}{Viktoria Vomhoff}, \bibinfo{person}{Michael Krolikowski}, \bibinfo{person}{Marco Mellia}, \bibinfo{person}{Diego Perino}, {and} \bibinfo{person}{Tobias Ho{\ss}feld}.} \bibinfo{year}{2024}\natexlab{}.
\newblock \showarticletitle{Untangling IoT Global Connectivity: The Importance of Mobile Signaling Traffic}.
\newblock \bibinfo{journal}{\emph{IEEE Transactions on Network and Service Management}} (\bibinfo{year}{2024}).
\newblock


\bibitem[Google({[n.\,d.]})]%
        {GoogleFi}
\bibfield{author}{\bibinfo{person}{Google}.} \bibinfo{year}{[n.\,d.]}\natexlab{}.
\newblock \bibinfo{title}{{ Google Fi Wireless }}.
\newblock \bibinfo{howpublished}{https://fi.google.com/about?pli=1}.
\newblock


\bibitem[Holafly({[n.\,d.]})]%
        {holafly}
\bibfield{author}{\bibinfo{person}{Holafly}.} \bibinfo{year}{[n.\,d.]}\natexlab{}.
\newblock \bibinfo{title}{{Stay connected wherever you go}}.
\newblock \bibinfo{howpublished}{https://esim.holafly.com/}.
\newblock


\bibitem[Jain et~al\mbox{.}(2022)]%
        {jain22L25GC}
\bibfield{author}{\bibinfo{person}{Vivek Jain}, \bibinfo{person}{Hao-Tse Chu}, \bibinfo{person}{Shixiong Qi}, \bibinfo{person}{Chia-An Lee}, \bibinfo{person}{Hung-Cheng Chang}, \bibinfo{person}{Cheng-Ying Hsieh}, \bibinfo{person}{K.~K. Ramakrishnan}, {and} \bibinfo{person}{Jyh-Cheng Chen}.} \bibinfo{year}{2022}\natexlab{}.
\newblock \showarticletitle{L25GC: a low latency 5G core network based on high-performance NFV platforms}. In \bibinfo{booktitle}{\emph{Proceedings of the ACM SIGCOMM 2022 Conference}} (Amsterdam, Netherlands) \emph{(\bibinfo{series}{SIGCOMM '22})}. \bibinfo{publisher}{Association for Computing Machinery}, \bibinfo{address}{New York, NY, USA}, \bibinfo{pages}{143–157}.
\newblock
\showISBNx{9781450394208}
\urldef\tempurl%
\url{https://doi.org/10.1145/3544216.3544267}
\showDOI{\tempurl}


\bibitem[Jang et~al\mbox{.}(2024)]%
        {jang2024unravelingairaloecosystem}
\bibfield{author}{\bibinfo{person}{Hyunseok~Daniel Jang}, \bibinfo{person}{Matteo Varvello}, \bibinfo{person}{Andra Lutu}, {and} \bibinfo{person}{Yasir Zaki}.} \bibinfo{year}{2024}\natexlab{}.
\newblock \bibinfo{title}{Unraveling the Airalo Ecosystem}.
\newblock
\newblock
\showeprint[arxiv]{2408.14923}~[cs.NI]
\urldef\tempurl%
\url{https://arxiv.org/abs/2408.14923}
\showURL{%
\tempurl}


\bibitem[Khan et~al\mbox{.}(2019)]%
        {khan2019survey}
\bibfield{author}{\bibinfo{person}{Rabia Khan}, \bibinfo{person}{Pardeep Kumar}, \bibinfo{person}{Dushantha Nalin~K Jayakody}, {and} \bibinfo{person}{Madhusanka Liyanage}.} \bibinfo{year}{2019}\natexlab{}.
\newblock \showarticletitle{A survey on security and privacy of 5G technologies: Potential solutions, recent advancements, and future directions}.
\newblock \bibinfo{journal}{\emph{IEEE Communications Surveys \& Tutorials}} \bibinfo{volume}{22}, \bibinfo{number}{1} (\bibinfo{year}{2019}), \bibinfo{pages}{196--248}.
\newblock


\bibitem[Klaa{\ss}en and Stoll(2021)]%
        {klaassen2021harmonizing}
\bibfield{author}{\bibinfo{person}{Lena Klaa{\ss}en} {and} \bibinfo{person}{Christian Stoll}.} \bibinfo{year}{2021}\natexlab{}.
\newblock \showarticletitle{Harmonizing corporate carbon footprints}.
\newblock \bibinfo{journal}{\emph{Nature communications}} \bibinfo{volume}{12}, \bibinfo{number}{1} (\bibinfo{year}{2021}), \bibinfo{pages}{1--13}.
\newblock


\bibitem[KORE({[n.\,d.]})]%
        {KORE}
\bibfield{author}{\bibinfo{person}{KORE}.} \bibinfo{year}{[n.\,d.]}\natexlab{}.
\newblock \bibinfo{title}{Harness the Power of 5G and IoT with KORE}.
\newblock \bibinfo{howpublished}{{ https://www.korewireless.com/ }}.
\newblock


\bibitem[Larrea et~al\mbox{.}(2023)]%
        {larrea23coreKube}
\bibfield{author}{\bibinfo{person}{Jon Larrea}, \bibinfo{person}{Andrew~E. Ferguson}, {and} \bibinfo{person}{Mahesh~K. Marina}.} \bibinfo{year}{2023}\natexlab{}.
\newblock \showarticletitle{CoreKube: An Efficient, Autoscaling and Resilient Mobile Core System}. In \bibinfo{booktitle}{\emph{Proceedings of the 29th Annual International Conference on Mobile Computing and Networking}} (Madrid, Spain) \emph{(\bibinfo{series}{ACM MobiCom '23})}. \bibinfo{publisher}{Association for Computing Machinery}, \bibinfo{address}{New York, NY, USA}, Article \bibinfo{articleno}{25}, \bibinfo{numpages}{15}~pages.
\newblock
\showISBNx{9781450399906}
\urldef\tempurl%
\url{https://doi.org/10.1145/3570361.3592522}
\showDOI{\tempurl}


\bibitem[Li et~al\mbox{.}(2018)]%
        {li2018measurement}
\bibfield{author}{\bibinfo{person}{Li Li}, \bibinfo{person}{Ke Xu}, \bibinfo{person}{Tong Li}, \bibinfo{person}{Kai Zheng}, \bibinfo{person}{Chunyi Peng}, \bibinfo{person}{Dan Wang}, \bibinfo{person}{Xiangxiang Wang}, \bibinfo{person}{Meng Shen}, {and} \bibinfo{person}{Rashid Mijumbi}.} \bibinfo{year}{2018}\natexlab{}.
\newblock \showarticletitle{A measurement study on multi-path TCP with multiple cellular carriers on high speed rails}. In \bibinfo{booktitle}{\emph{Proceedings of the 2018 Conference of the ACM Special Interest Group on Data Communication}}. \bibinfo{pages}{161--175}.
\newblock


\bibitem[Liyanage et~al\mbox{.}(2016)]%
        {liyanage2016opportunities}
\bibfield{author}{\bibinfo{person}{Madhusanka Liyanage}, \bibinfo{person}{Ahmed~Bux Abro}, \bibinfo{person}{Mika Ylianttila}, {and} \bibinfo{person}{Andrei Gurtov}.} \bibinfo{year}{2016}\natexlab{}.
\newblock \showarticletitle{Opportunities and challenges of software-defined mobile networks in network security}.
\newblock \bibinfo{journal}{\emph{IEEE security \& privacy}} \bibinfo{volume}{14}, \bibinfo{number}{4} (\bibinfo{year}{2016}), \bibinfo{pages}{34--44}.
\newblock


\bibitem[Luo et~al\mbox{.}(2021)]%
        {luo21cellbricks}
\bibfield{author}{\bibinfo{person}{Zhihong Luo}, \bibinfo{person}{Silvery Fu}, \bibinfo{person}{Mark Theis}, \bibinfo{person}{Shaddi Hasan}, \bibinfo{person}{Sylvia Ratnasamy}, {and} \bibinfo{person}{Scott Shenker}.} \bibinfo{year}{2021}\natexlab{}.
\newblock \showarticletitle{Democratizing cellular access with CellBricks}. In \bibinfo{booktitle}{\emph{Proceedings of the 2021 ACM SIGCOMM 2021 Conference}} (Virtual Event, USA) \emph{(\bibinfo{series}{SIGCOMM '21})}. \bibinfo{publisher}{Association for Computing Machinery}, \bibinfo{address}{New York, NY, USA}, \bibinfo{pages}{626–640}.
\newblock
\showISBNx{9781450383837}
\urldef\tempurl%
\url{https://doi.org/10.1145/3452296.3473336}
\showDOI{\tempurl}


\bibitem[Lutu et~al\mbox{.}(2020a)]%
        {lutu2020dice}
\bibfield{author}{\bibinfo{person}{Andra Lutu}, \bibinfo{person}{Marcelo Bagnulo}, {and} \bibinfo{person}{Diego Perino}.} \bibinfo{year}{2020}\natexlab{a}.
\newblock \bibinfo{title}{DICE: Dynamic Interconnections for the Cellular Ecosystem}.
\newblock
\newblock
\showeprint[arxiv]{2007.13591}~[cs.NI]
\urldef\tempurl%
\url{https://arxiv.org/abs/2007.13591}
\showURL{%
\tempurl}


\bibitem[Lutu et~al\mbox{.}(2020b)]%
        {lutu2020first}
\bibfield{author}{\bibinfo{person}{Andra Lutu}, \bibinfo{person}{Byungjin Jun}, \bibinfo{person}{Fabi{\'a}n~E Bustamante}, \bibinfo{person}{Diego Perino}, \bibinfo{person}{Marcelo Bagnulo}, {and} \bibinfo{person}{Carlos~Gamboa Bontje}.} \bibinfo{year}{2020}\natexlab{b}.
\newblock \showarticletitle{A first look at the ip exchange ecosystem}.
\newblock \bibinfo{journal}{\emph{ACM SIGCOMM Computer Communication Review}} \bibinfo{volume}{50}, \bibinfo{number}{4} (\bibinfo{year}{2020}), \bibinfo{pages}{25--34}.
\newblock


\bibitem[Lutu et~al\mbox{.}(2020c)]%
        {lutu2020things}
\bibfield{author}{\bibinfo{person}{Andra Lutu}, \bibinfo{person}{Byungjin Jun}, \bibinfo{person}{Alessandro Finamore}, \bibinfo{person}{Fabi{\'a}n~E Bustamante}, {and} \bibinfo{person}{Diego Perino}.} \bibinfo{year}{2020}\natexlab{c}.
\newblock \showarticletitle{Where Things Roam: Uncovering Cellular IoT/M2M Connectivity}. In \bibinfo{booktitle}{\emph{Proceedings of the ACM Internet Measurement Conference}}. \bibinfo{pages}{147--161}.
\newblock


\bibitem[Lutu et~al\mbox{.}(2021)]%
        {lutu2021insights}
\bibfield{author}{\bibinfo{person}{Andra Lutu}, \bibinfo{person}{Diego Perino}, \bibinfo{person}{Marcelo Bagnulo}, {and} \bibinfo{person}{Fabi{\'a}n~E Bustamante}.} \bibinfo{year}{2021}\natexlab{}.
\newblock \showarticletitle{Insights from operating an IP exchange provider}. In \bibinfo{booktitle}{\emph{Proceedings of the 2021 ACM SIGCOMM 2021 Conference}}. \bibinfo{pages}{718--730}.
\newblock


\bibitem[Macia-Fernandez et~al\mbox{.}(2009)]%
        {macia2009fraud}
\bibfield{author}{\bibinfo{person}{Gabriel Macia-Fernandez}, \bibinfo{person}{Pedro Garcia-Teodoro}, {and} \bibinfo{person}{Jesus Diaz-Verdejo}.} \bibinfo{year}{2009}\natexlab{}.
\newblock \showarticletitle{Fraud in roaming scenarios: An overview}.
\newblock \bibinfo{journal}{\emph{IEEE Wireless Communications}} \bibinfo{volume}{16}, \bibinfo{number}{6} (\bibinfo{year}{2009}), \bibinfo{pages}{88--94}.
\newblock


\bibitem[Mafakheri et~al\mbox{.}(2021)]%
        {mafakheri2021smart}
\bibfield{author}{\bibinfo{person}{Babak Mafakheri}, \bibinfo{person}{Andreas Heider-Aviet}, \bibinfo{person}{Roberto Riggio}, {and} \bibinfo{person}{Leonardo Goratti}.} \bibinfo{year}{2021}\natexlab{}.
\newblock \showarticletitle{Smart contracts in the 5G roaming architecture: the fusion of blockchain with 5G networks}.
\newblock \bibinfo{journal}{\emph{IEEE Communications Magazine}} \bibinfo{volume}{59}, \bibinfo{number}{3} (\bibinfo{year}{2021}), \bibinfo{pages}{77--83}.
\newblock


\bibitem[Mahyoub et~al\mbox{.}(2024)]%
        {mahyoub2024security}
\bibfield{author}{\bibinfo{person}{Mohammed Mahyoub}, \bibinfo{person}{AbdulAziz AbdulGhaffar}, \bibinfo{person}{Emmanuel Alalade}, \bibinfo{person}{Ezekiel Ndubisi}, {and} \bibinfo{person}{Ashraf Matrawy}.} \bibinfo{year}{2024}\natexlab{}.
\newblock \showarticletitle{Security analysis of critical 5g interfaces}.
\newblock \bibinfo{journal}{\emph{IEEE Communications Surveys \& Tutorials}} (\bibinfo{year}{2024}).
\newblock


\bibitem[Mandalari et~al\mbox{.}(2022)]%
        {mandalari2022measuring}
\bibfield{author}{\bibinfo{person}{Anna~Maria Mandalari}, \bibinfo{person}{Andra Lutu}, \bibinfo{person}{Ana Custura}, \bibinfo{person}{Ali~Safari Khatouni}, \bibinfo{person}{Özgü Alay}, \bibinfo{person}{Marcelo Bagnulo}, \bibinfo{person}{Vaibhav Bajpai}, \bibinfo{person}{Anna Brunstrom}, \bibinfo{person}{Jörg Ott}, \bibinfo{person}{Martino Trevisan}, \bibinfo{person}{Marco Mellia}, {and} \bibinfo{person}{Gorry Fairhurst}.} \bibinfo{year}{2022}\natexlab{}.
\newblock \showarticletitle{Measuring Roaming in Europe: Infrastructure and Implications on Users’ QoE}.
\newblock \bibinfo{journal}{\emph{IEEE Transactions on Mobile Computing}} \bibinfo{volume}{21}, \bibinfo{number}{10} (\bibinfo{year}{2022}), \bibinfo{pages}{3687--3699}.
\newblock
\urldef\tempurl%
\url{https://doi.org/10.1109/TMC.2021.3058787}
\showDOI{\tempurl}


\bibitem[Mandalari et~al\mbox{.}(2018)]%
        {mandalari2018experience}
\bibfield{author}{\bibinfo{person}{Anna~Maria Mandalari}, \bibinfo{person}{Andra Lutu}, \bibinfo{person}{Ana Custura}, \bibinfo{person}{Ali Safari~Khatouni}, \bibinfo{person}{{\"O}zg{\"u} Alay}, \bibinfo{person}{Marcelo Bagnulo}, \bibinfo{person}{Vaibhav Bajpai}, \bibinfo{person}{Anna Brunstrom}, \bibinfo{person}{J{\"o}rg Ott}, \bibinfo{person}{Marco Mellia}, {et~al\mbox{.}}} \bibinfo{year}{2018}\natexlab{}.
\newblock \showarticletitle{Experience: Implications of roaming in europe}. In \bibinfo{booktitle}{\emph{Proceedings of the 24th Annual International Conference on Mobile Computing and Networking}}. \bibinfo{pages}{179--189}.
\newblock


\bibitem[Mobile({[n.\,d.]})]%
        {yoho}
\bibfield{author}{\bibinfo{person}{Yoho Mobile}.} \bibinfo{year}{[n.\,d.]}\natexlab{}.
\newblock \bibinfo{title}{{Stay Connected Anytime and Anywhere}}.
\newblock \bibinfo{howpublished}{https://yohomobile.com/}.
\newblock


\bibitem[Nomad({[n.\,d.]})]%
        {nomad}
\bibfield{author}{\bibinfo{person}{Nomad}.} \bibinfo{year}{[n.\,d.]}\natexlab{}.
\newblock \bibinfo{title}{{Nomad eSIM: Prepaid Data Plan}}.
\newblock \bibinfo{howpublished}{https://www.getnomad.app/}.
\newblock


\bibitem[onesim({[n.\,d.]})]%
        {onesim}
\bibfield{author}{\bibinfo{person}{onesim}.} \bibinfo{year}{[n.\,d.]}\natexlab{}.
\newblock \bibinfo{title}{{Best prepaid eSIM data plans for travelers, digital nomads, and remote workers}}.
\newblock \bibinfo{howpublished}{https://onesim.co/}.
\newblock


\bibitem[Prasad and Sridhar(2023)]%
        {prasad20235g}
\bibfield{author}{\bibinfo{person}{Rohit Prasad} {and} \bibinfo{person}{V Sridhar}.} \bibinfo{year}{2023}\natexlab{}.
\newblock \bibinfo{booktitle}{\emph{5G and Beyond: Formulating a Regulatory Response}}.
\newblock \bibinfo{type}{{T}echnical {R}eport}. \bibinfo{institution}{Indian Council for Research on International Economic Relations (ICRIER~…}.
\newblock


\bibitem[Refaey et~al\mbox{.}(2019)]%
        {refaey2019blockchain}
\bibfield{author}{\bibinfo{person}{Ahmed Refaey}, \bibinfo{person}{Karim Hammad}, \bibinfo{person}{Sebastian Magierowski}, {and} \bibinfo{person}{Ekram Hossain}.} \bibinfo{year}{2019}\natexlab{}.
\newblock \showarticletitle{A blockchain policy and charging control framework for roaming in cellular networks}.
\newblock \bibinfo{journal}{\emph{IEEE Network}} \bibinfo{volume}{34}, \bibinfo{number}{3} (\bibinfo{year}{2019}), \bibinfo{pages}{170--177}.
\newblock


\bibitem[Romero~Ugarte(2018)]%
        {romero2018distributed}
\bibfield{author}{\bibinfo{person}{Jos{\'e}~Luis Romero~Ugarte}.} \bibinfo{year}{2018}\natexlab{}.
\newblock \showarticletitle{Distributed ledger technology (DLT): introduction}.
\newblock \bibinfo{journal}{\emph{Banco de Espana Article}}  \bibinfo{volume}{19} (\bibinfo{year}{2018}), \bibinfo{pages}{18}.
\newblock


\bibitem[Sadhukhan et~al\mbox{.}(2024)]%
        {sadhukhan2024development}
\bibfield{author}{\bibinfo{person}{Dipanwita Sadhukhan}, \bibinfo{person}{Sangram Ray}, \bibinfo{person}{Mou Dasgupta}, {and} \bibinfo{person}{Muhammad~Khurram Khan}.} \bibinfo{year}{2024}\natexlab{}.
\newblock \showarticletitle{Development of a provably secure and privacy-preserving lightweight authentication scheme for roaming services in global mobility network}.
\newblock \bibinfo{journal}{\emph{Journal of Network and Computer Applications}}  \bibinfo{volume}{224} (\bibinfo{year}{2024}), \bibinfo{pages}{103831}.
\newblock


\bibitem[Schmitt et~al\mbox{.}(2022)]%
        {schmitt2022decoupling}
\bibfield{author}{\bibinfo{person}{Paul Schmitt}, \bibinfo{person}{Jana Iyengar}, \bibinfo{person}{Christopher Wood}, {and} \bibinfo{person}{Barath Raghavan}.} \bibinfo{year}{2022}\natexlab{}.
\newblock \showarticletitle{The decoupling principle: a practical privacy framework}. In \bibinfo{booktitle}{\emph{Proceedings of the 21st ACM Workshop on Hot Topics in Networks}}. \bibinfo{pages}{213--220}.
\newblock


\bibitem[Schmitt and Raghavan(2021)]%
        {schmitt2021pretty}
\bibfield{author}{\bibinfo{person}{Paul Schmitt} {and} \bibinfo{person}{Barath Raghavan}.} \bibinfo{year}{2021}\natexlab{}.
\newblock \showarticletitle{Pretty good phone privacy}. In \bibinfo{booktitle}{\emph{30th USENIX Security Symposium (USENIX Security 21)}}. \bibinfo{pages}{1737--1754}.
\newblock


\bibitem[Schmitt et~al\mbox{.}(2016a)]%
        {schmitt2016study}
\bibfield{author}{\bibinfo{person}{Paul Schmitt}, \bibinfo{person}{Morgan Vigil}, {and} \bibinfo{person}{Elizabeth Belding}.} \bibinfo{year}{2016}\natexlab{a}.
\newblock \showarticletitle{A study of MVNO data paths and performance}. In \bibinfo{booktitle}{\emph{Passive and Active Measurement: 17th International Conference, PAM 2016, Heraklion, Greece, March 31-April 1, 2016. Proceedings 17}}. Springer, \bibinfo{pages}{83--94}.
\newblock


\bibitem[Schmitt et~al\mbox{.}(2016b)]%
        {schmitt16pam}
\bibfield{author}{\bibinfo{person}{Paul Schmitt}, \bibinfo{person}{Morgan Vigil}, {and} \bibinfo{person}{Elizabeth Belding}.} \bibinfo{year}{2016}\natexlab{b}.
\newblock \showarticletitle{A study of MVNO data paths and performance}. In \bibinfo{booktitle}{\emph{International Conference on Passive and Active Network Measurement}}. Springer, \bibinfo{pages}{83--94}.
\newblock


\bibitem[Twilio({[n.\,d.]})]%
        {twilio}
\bibfield{author}{\bibinfo{person}{Twilio}.} \bibinfo{year}{[n.\,d.]}\natexlab{}.
\newblock \bibinfo{title}{The trusted platform for data-driven customer engagement across any channel.}
\newblock \bibinfo{howpublished}{{ https://www.twilio.com/}}.
\newblock


\bibitem[Voigt and Von~dem Bussche(2017)]%
        {voigt2017eu}
\bibfield{author}{\bibinfo{person}{Paul Voigt} {and} \bibinfo{person}{Axel Von~dem Bussche}.} \bibinfo{year}{2017}\natexlab{}.
\newblock \showarticletitle{The eu general data protection regulation (gdpr)}.
\newblock \bibinfo{journal}{\emph{A Practical Guide, 1st Ed., Cham: Springer International Publishing}} \bibinfo{volume}{10}, \bibinfo{number}{3152676} (\bibinfo{year}{2017}), \bibinfo{pages}{10--5555}.
\newblock


\bibitem[Vomhoff et~al\mbox{.}(2024a)]%
        {vomhoff2024shortcut}
\bibfield{author}{\bibinfo{person}{Viktoria Vomhoff}, \bibinfo{person}{Marleen Sichermann}, \bibinfo{person}{Stefan Giei{\ss}ler}, \bibinfo{person}{Andra Lutu}, \bibinfo{person}{Martin Giess}, {and} \bibinfo{person}{Tobias Ho{\ss}feld}.} \bibinfo{year}{2024}\natexlab{a}.
\newblock \showarticletitle{A Shortcut Through the IPX: Measuring Latencies in Global Mobile Roaming with Regional Breakouts}. In \bibinfo{booktitle}{\emph{2024 8th Network Traffic Measurement and Analysis Conference (TMA)}}. IEEE, \bibinfo{pages}{1--10}.
\newblock


\bibitem[Vomhoff et~al\mbox{.}(2024b)]%
        {vomhoffshortcut}
\bibfield{author}{\bibinfo{person}{Viktoria Vomhoff}, \bibinfo{person}{Marleen Sichermann}, \bibinfo{person}{Stefan Gieißler}, \bibinfo{person}{Andra Lutu}, \bibinfo{person}{Martin Giess}, {and} \bibinfo{person}{Tobias Hoßfeld}.} \bibinfo{year}{2024}\natexlab{b}.
\newblock \showarticletitle{A Shortcut Through the IPX: Measuring Latencies in Global Mobile Roaming with Regional Breakouts}. In \bibinfo{booktitle}{\emph{2024 8th Network Traffic Measurement and Analysis Conference (TMA)}}. \bibinfo{pages}{1--10}.
\newblock


\bibitem[Xiao et~al\mbox{.}(2019)]%
        {xiao19mobisys}
\bibfield{author}{\bibinfo{person}{Ao Xiao}, \bibinfo{person}{Yunhao Liu}, \bibinfo{person}{Yang Li}, \bibinfo{person}{Feng Qian}, \bibinfo{person}{Zhenhua Li}, \bibinfo{person}{Sen Bai}, \bibinfo{person}{Yao Liu}, \bibinfo{person}{Tianyin Xu}, {and} \bibinfo{person}{Xianlong Xin}.} \bibinfo{year}{2019}\natexlab{}.
\newblock \showarticletitle{An in-depth study of commercial MVNO: Measurement and optimization}. In \bibinfo{booktitle}{\emph{Proceedings of the 17th Annual International Conference on Mobile Systems, Applications, and Services}} (Seoul, Republic of Korea) \emph{(\bibinfo{series}{MobiSys '19})}. \bibinfo{pages}{457--468}.
\newblock


\bibitem[Xie et~al\mbox{.}(2014)]%
        {xie2014privacy}
\bibfield{author}{\bibinfo{person}{Qi Xie}, \bibinfo{person}{Dongzhao Hong}, \bibinfo{person}{Mengjie Bao}, \bibinfo{person}{Na Dong}, {and} \bibinfo{person}{Duncan~S Wong}.} \bibinfo{year}{2014}\natexlab{}.
\newblock \showarticletitle{Privacy-preserving mobile roaming authentication with security proof in global mobility networks}.
\newblock \bibinfo{journal}{\emph{International Journal of Distributed Sensor Networks}} \bibinfo{volume}{10}, \bibinfo{number}{5} (\bibinfo{year}{2014}), \bibinfo{pages}{325734}.
\newblock


\bibitem[yesim({[n.\,d.]})]%
        {yesim}
\bibfield{author}{\bibinfo{person}{yesim}.} \bibinfo{year}{[n.\,d.]}\natexlab{}.
\newblock \bibinfo{title}{{eSIM Mobile Data Plans for Tourism and Business}}.
\newblock \bibinfo{howpublished}{https://yesim.app/}.
\newblock


\bibitem[Yuan et~al\mbox{.}(2018)]%
        {yuan18mobicom}
\bibfield{author}{\bibinfo{person}{Zengwen Yuan}, \bibinfo{person}{Qianru Li}, \bibinfo{person}{Yuanjie Li}, \bibinfo{person}{Songwu Lu}, \bibinfo{person}{Chunyi Peng}, {and} \bibinfo{person}{George Varghese}.} \bibinfo{year}{2018}\natexlab{}.
\newblock \showarticletitle{Resolving Policy Conflicts in Multi-Carrier Cellular Access}. In \bibinfo{booktitle}{\emph{Proceedings of the 24th Annual International Conference on Mobile Computing and Networking}} (New Delhi, India) \emph{(\bibinfo{series}{MobiCom '18})}. \bibinfo{publisher}{Association for Computing Machinery}, \bibinfo{address}{New York, NY, USA}, \bibinfo{pages}{147--162}.
\newblock
\showISBNx{9781450359030}
\urldef\tempurl%
\url{https://doi.org/10.1145/3241539.3241558}
\showDOI{\tempurl}


\bibitem[Zarinni et~al\mbox{.}(2014a)]%
        {zarinni2014first}
\bibfield{author}{\bibinfo{person}{Fatima Zarinni}, \bibinfo{person}{Ayon Chakraborty}, \bibinfo{person}{Vyas Sekar}, \bibinfo{person}{Samir~R Das}, {and} \bibinfo{person}{Phillipa Gill}.} \bibinfo{year}{2014}\natexlab{a}.
\newblock \showarticletitle{A first look at performance in mobile virtual network operators}. In \bibinfo{booktitle}{\emph{Proceedings of the 2014 conference on internet measurement conference}}. \bibinfo{pages}{165--172}.
\newblock


\bibitem[Zarinni et~al\mbox{.}(2014b)]%
        {zarinni14imc}
\bibfield{author}{\bibinfo{person}{Fatima Zarinni}, \bibinfo{person}{Ayon Chakraborty}, \bibinfo{person}{Vyas Sekar}, \bibinfo{person}{Samir~R Das}, {and} \bibinfo{person}{Phillipa Gill}.} \bibinfo{year}{2014}\natexlab{b}.
\newblock \showarticletitle{A first look at performance in mobile virtual network operators}. In \bibinfo{booktitle}{\emph{Proceedings of the 2014 conference on internet measurement conference}} (Vancouver, BC, Canada) \emph{(\bibinfo{series}{IMC '14})}. \bibinfo{pages}{165--172}.
\newblock


\end{thebibliography}

\appendix
\printglossaries

\end{document}